\documentclass{css2002}

\begin{document}

\title{Investigation of light scalar resonances at COSY}

\author{M.B\"uscher, F.P.Sassen
\address{Institut f\"ur Kernphysik, Forschungszentrum J\"ulich, 
           52425 J\"ulich, Germany}
\and
N.N.Achasov,
\address{Laboratory of Theoretical Physics, Sobolev Institute for Mathematics,
            630090 Novosibirsk, Russia}
\and
L.Kondratyuk
\address{Institute of Theoretical and Experimental physics, 
         B.~Cheremushkinskaya 25, 117259 Moscow, Russia} 
%
 %Shortcut of the title for topline of odd pages
 \headtitle{Light scalar resonances}
 %up to three authors on topline of even pages
 \headauthor{M.B\"uscher, F.P.Sassen, N.N.Achasov et al.}  }

\maketitle 
\begin{abstract}
  The $a_0$(980) and $f_0$(980) resonances are two well established
  states in the excited meson spectrum. We review the most prominent
  theoretical models which try to explain the structure of these
  states. It is discussed whether data from COSY on $a_0$ and $f_0$
  production in $pp$, $pn$, $pd$ and $dd$ collisions allow to
  distinguish between the different approaches. Very promising in this
  respect seems to be the measurement of the reaction $dd \rightarrow
  (\mathrm{^4He}\, a_0^0 \rightarrow) \,\mathrm{^4He}\, \pi^0\eta$
  which violates isospin conservation and can be related to
  $a_0$-$f_0$ mixing.
\end{abstract}

\section{Overview of theoretical models}
So far a final consensus on the nature of the light scalar mesons has
not been reached. In contrast to the case of the $\sigma$ where new
data on the in-medium modifications \cite{TAPS} support the picture of
two correlated pions the subject remains quite open concerning the
nature of the $a_0$ and $f_0$. In the following a short list of some
interpretations given.
\begin{description}
\item[$q\bar{q}$-state:] Despite the very successful mass predictions,
  e.g.\ for vector mesons, coming from relativistic constituent quark
  models, mass predictions in the scalar sector still vary
  considerably. The predictions depend strongly on the choice for the
  Dirac structure of the confining potential and the models cannot
  explain the reason for the $a_0$/$f_0$-mass degeneracy. In
  Ref.~\cite{RRicken} either the mass of the $a_0$ or the $f_0$ is
  reproduced depending on the spin parametrisation of confinement.
  Other authors \cite{LSCelenza} deduce from their relativistic quark
  model the need for a glueball mixing with the calculated states to
  reproduce experimental data. Within those mixing schemes
  interpretations vary from minimal gluonic admixture to the
  $q\bar{q}$-states $f_0$ and $a_0$ \cite{VVAnisovich} up to large
  gluonic components \cite{SNarison}. Thus the answers coming from
  $q\bar{q}$-models are not conclusive yet.
\item[\boldmath$(qq)(\bar{q}\bar{q})$-states:] This structure allows
  for two configurations in colour space: $\{\bar{3}3\}$ and
  $\{6\bar{6}\}$.  Of course those two configurations may mix/
  rearrange to form a state like $(q\bar{q})(q\bar{q})$ with colour
  configuration $\{11\}$.  Therefore, the number of expected
  tetraquark states and their interpretation vary for different
  calculations and it is difficult to distinguish a mesonic molecule
  from a tetraquark state.  In Ref.~\cite{JVijande} for example 5
  tetraquarks are found. From those two are nearly mass degenerate and
  obtain the mass of the $a_0$ and $f_0$ when a fudge factor is used
  to account for neglected three and four body forces. This opposes
  the complete tetraquark nonet expected in Ref.~\cite{FEClose}, which
  might result from the authors considering pure
  $(qq)_{\bar{3}}(\bar{q}\bar{q})_3$ and only allowing
  $(q\bar{q})_1(q\bar{q})_1$ admixture for large distances, whereas
  this component might occur at small distances by mixing of
  $\{\bar{3}3\}$, $\{6\bar{6}\}$ configurations, resulting in a less
  attractive interaction. This was for example pointed out in
  Ref.~\cite{JWeinstein}. In the latter paper it was observed, that
  the four quarks in the confining potential are mainly arranged as
  two colour singlets at large $(1.5$ fm$)$ distance and the notion of
  a mesonic molecule was introduced.
\item[Mesonic molecules:] Inspired by the original mesonic molecule in
  Ref.~\cite{JWeinstein} (see the $(qq)(\bar{q}\bar{q})$ case) a
  second molecule picture \cite{DLohse} was developed in which only
  mesonic degrees of freedom (i.e.\ colour singlets) were considered.
  Since $\rho$-exchange between the $K\bar{K}$-pair sets the scale for
  the bound state identified as $f_0$, it might be even more compact
  than the molecule of Ref.~\cite{JWeinstein}.  This is why radiative
  $\phi$ decays exclude the extended molecule of
  Ref.~\cite{JWeinstein} while still giving strong evidence in favour
  of a compact $K\bar{K}$-state or, as it is phrased in
  Ref.~\cite{NNAchasov}, a compact four-quark state. Furthermore the
  chiral unitary approach, which in its structure is very close to the
  mesonic model, has no problems to describe radiative $\phi$-decays
  \cite{JAOller}, which is a hint that the discussed $K\bar{K}$ state
  really is compact.
\item[Chiral unitary approach:] Inspired by chiral perturbation
  theory, schemes have been developed to deal with unitarity. As above
  those methods generate bound states from two mesons \cite{JAOller2}.
  These states differ from mesonic molecules only by the theoretical
  framework used to calculate them and not by physical content.
\item[Semi-bound states:] In some models the attraction for the
  mesonic sector is not sufficient to generate bound states.  One
  example for this is the cusp-effect in Ref.~\cite{DLohse} which is
  identified with a $a_0$ signal.  Another example would be the
  model of Ref.~\cite{EvanBeveren}. Here the light scalar mesons do
  not stem from the confining potential but they have their origin in
  the $^3P_0$-barrier, which is just too low to generate bound
  meson-meson states.
\item[Mixing-schemes:] A different approach to explain the abundance of
  observed scalar states is to introduce a scalar glueball around
  $1.4$--$1.8$ GeV as predicted by lattice calculations. This glueball
  then mixes with the other scalar states to form the observed
  resonances. Depending on the predicted masses of the glueball and
  the bare states different mixings are obtained, e.g.\ \cite{AKirk}.
\item[Vacuum scalars:] This idea of identifying $f_0$ and $a_0$ as
  systems formed of negative kinetic energy $u$ and $d$ flavours
  \cite{VNGribov} seems to be ruled out by experiment even though the
  predictive power of this model was plagued by the
  complexity of the calculations. The predicted compact size of the
  object contrasts too much with the decay constant ratio of $a_0$ and
  $K_0^*$ as shown in \cite{KMaltman}. 
\end{description}
During the discussion in the workshop, tetraquark models were favoured
whereas no conclusion was drawn on the characteristic of the different
four-quark states. No key observables have been identified to
discriminate mesonic molecules, chiral unitarity effects and
$(qq)(\bar{q}\bar{q})$-states. It was pointed out that the available
data set on the production of the light scalars in hadronic
interactions needs to be extended. Measurements of as many observables
as possible with different beam($p,\vec{p},d$)-target($p,n,d$)
combinations and tests of isospin ratios should be performed. In
particular, the observation of the process $dd\to\, ^4\mathrm{He}
\pi^0\eta$ which is forbidden by isospin conservation might yield
information about the strength of the $a_0^0$-$f_0$ mixing.

\section{Available data from COSY}
Data about the production and decay of the light scalar resonances
from COSY are rather scarce yet. Some information about the
$a_0^0/f_0$-production cross sections in $pp$ and $pd$ reactions can
be deduced model dependently \cite{Brat01} from data on $K^+K^-$
production. Such measurements have been performed at the COSY-11 and
MOMO facilities at beam energies close to the $K^+K^-$-production
threshold, see Table~\ref{tab:data}. 

\begin{table}[htbp]
  \begin{center}
    \begin{tabular}[t]{r|c|c|c|c}
Experiment & Reaction & $Q$ & $\sigma_{\mathrm{tot}}$ & contribution  \\
           &          &(MeV)&  (nb)                   & via $a_0^0/f_0$\\
\hline
&&&&\\
COSY-11  \cite{COSY11}   & $pp \to pp\,K^+K^-$ & 17 & $1.8\pm0.27^{+0.26}_{-0.35}$ &small \cite{Brat01}\\
&&&&\\
MOMO \cite{MOMO}         & $pd \to\, ^3\mathrm{He}\,K^+K^-$ & 40 & $9.6\pm1.0$ &?\\
&& 56 & $17.5\pm1.8$  & ? \\
&&&&\\
ANKE                     & $pp \to\, d\,K^+\bar{K}^0$ & 44 & $45\pm6\pm16$ &$\sim70$\% \cite{MESON2002}\\
&&&{\em (preliminary)}&\\  
&&&&\\
    \end{tabular}
    \caption{Overview over the available data from COSY}
    \label{tab:data}
  \end{center}
\end{table}

According to the model calculations outlined in Ref.~\cite{Brat01},
the fraction of resonantly produced $K\bar{K}$ pairs (via the
$a_0^0/f_0$) in $pp \to pn\,K^+\bar{K}^0$ reactions is significantly
larger than for the $pp \to pp\,K^+K^-$ case. Following this idea, two
measurements of the reactions $pp \to d\,K^+\bar{K}^0/\pi^+\eta$ have
been performed at the ANKE spectrometer for $Q=44$ and 104 MeV.  The
total $pp \to d\,K^+\bar{K}^0$ cross section at $Q{=}44$ MeV is about
$\sim$45~nb \cite{Fedorets_QNP2002}, in excelent agreement with the
model predictions ($\sim$40~nb). The same model predicts a resonant
contribution of $\sim$70\% for this $Q$ value \cite{MESON2002}. This
interpretation is in line with a preliminary analysis of the
$\pi^+\eta$ decay channel where a resonant structure around
980~MeV/c$^2$ is seen in the invariant $\pi^+\eta$-mass distribution
with a width of $\Gamma\sim 40$~MeV/c$^2$ \cite{Fedorets_QNP2002}.
From these data it can be concluded that systematic studies of the
light scalar mesons are possible with ANKE.

\section{Planned measurements}
It was suggested long ago that the coupling of the $a_0(980)$- and
$f_0(980)$-resonances to the $K \bar K$ continuum should give rise to
a significant $a_0^0$-$f_0$ mixing in the vicinity of the $K \bar K$
threshold \cite{Achasov}.  Different aspects of this mixing, the
underlying dynamics, and the possibilities to measure this effect have
been discussed in Refs.\
\cite{Krehl,Achasov2,Grishina2001,Kudryavtsev,Kudryavtsev2}.  It has
been suggested by Close and Kirk~\cite{Clo} that new data from the
WA102 collaboration at CERN~\cite{WA102} on the central production of
$a_0$ and $f_0$ in the reaction $pp\to p_{\mathrm s} X p_{\mathrm f}$
provide evidence for a significant $a_0$-$f_0$-mixing intensity
$|\xi|^2=8\pm 3$\%.

Possible experimental tests of isospin violation due to $a_0$-$f_0$
mixing based on a combined analysis of the reactions
\begin{eqnarray}
pp \to d\,a_0^+\ \ \ &\mathrm{and}&\ \ pn \to d\,a_0^0\ \\
pd \to\,^3\mathrm{H}\,a_0^+\ \ \ &\mathrm{and}&\ \  pd \to\,^3\mathrm{He}\,a_0^0
\end{eqnarray}
are discussed in Ref.~\cite{Grishina2001}. A corresponding proposal
for measurements at ANKE \cite{a0f0_proposal} has already been
approved by the COSY-PAC and the measurements are planned for winter
2003/04.

Direct production of the $a_0$ resonance in the reaction $dd
\rightarrow \mathrm{^4He}\, a_0^0$ is forbidden if isospin is
conserved.  It can, for example, be observed due to  $a_0$-$f_0$
mixing
\begin{equation}
  \sigma(dd \rightarrow \mathrm{^4He}\, a_0^0)
  = |\xi|^2\ \cdot \sigma(dd \rightarrow \mathrm{^4He}\, f_0).
  \label{eq:dd2hef0}
\end{equation}
Therefore it is very interesting to study the reaction
\begin{equation}
  dd \rightarrow \mathrm{^4He}\, (\pi^0~ \eta) \label{dd}
\end{equation}
at $m^2_{\pi\eta}\sim(980\, \mathrm{MeV})^2$. Any signal of
reaction (\ref{dd}) will be related to isospin breaking, which is
expected to be more pronounced near the $f_0$ threshold as compared to
the region below (or above).

An important point for the feasibility of such measurements is the
magnitude of the cross sections $\sigma (dd \rightarrow
\mathrm{^4He}\, a_0^0)$ and $\sigma (dd \rightarrow \mathrm{^4He}\,
f_0)$. Experimental data are not available yet and we try to give a
qualitative estimate of these cross sections: According to
Refs.~\cite{Frascaria,Willis,Zlomanchuk}, the cross-section ratio
$\sigma (dd \rightarrow \mathrm{^4He}\, \eta )/ \sigma (dd \rightarrow
\mathrm{^3He}\, \eta)$ is about 0.04 at $Q \simeq 10$ MeV. We assume
an approximately equal ratio for the case of $K^+ K^-$ production near
the threshold:
\begin{equation}
  \sigma(dd \rightarrow \mathrm{^4He}\, K^+ K^-) =0.04\cdot 
  \sigma(pd \rightarrow \mathrm{^3He}\, K^+ K^-)\ .
\end{equation} 

Using the MOMO data \cite{MOMO} on the reaction $pd\rightarrow
\mathrm{^3He}\, K^+ K^-$ (see Table~\ref{tab:data}) we find:
\begin{equation}
  \sigma(dd\rightarrow \mathrm{^4He}\, K^+ K^-)\simeq 0.4~\mathrm{nb}
  \label{ddalfaKK}
\end{equation} 
at $Q{=}40$ MeV.  The MOMO collaboration notes that their invariant
$K^+K^-$-mass distributions follow phase space.  However, as it was
shown for the case of the $a_0$ resonance in Ref.~\cite{Brat01}, the
shape of the invariant mass spectrum following phase space cannot be
distinguished from resonance production at $Q \leq \Gamma \leq 70$
MeV.

Therefore, the broad mass distribution of the MOMO data may also be
related to the $f_0$ (or $a_0$ ).  This statement is supported by a
two-step model where the amplitude of the reaction $pd \rightarrow
\mathrm{^3He}\, f_0$ can be constructed from the subprocesses $pp
\rightarrow d \pi^+$ and $\pi^+ n \rightarrow p\, f_0$ (cf.\ 
Refs.~\cite{Faldt,Uzikov}). As it is known from the available
experimental data \cite{Landolt} the cross section of the reaction
$\pi N \to N K \bar K$ near threshold has an essential contribution
from the $f_0$ resonance in the case of isoscalar $K \bar K$
production.  Thus the cross section of the reaction $pd \rightarrow
\mathrm{^3He}\,f_0 \rightarrow \mathrm{^3He}\,K^+K^-$ near threshold
is expected to be not significantly smaller than the upper limit from
MOMO of about $10\div 20$ nb at $Q=40\div60$ MeV.

For an estimate of $\sigma(dd\rightarrow \mathrm{^4He}\, \pi^+ \pi^-$)
at $m_{\pi\pi}\sim m_{f_o}$ we assume that the cross section
$\sigma(dd\rightarrow \mathrm{^4He}\, K^+ K^-)$ is also dominated by
resonant $f_0$ production at $m_{K\bar{K}}\sim m_{f_o}$ , and that
$\Gamma_{f_0\to K \bar K}=(0.1\div0.4) \cdot \Gamma_{f_0 \to \pi \pi}$
\cite{PDG}. This yields
\begin{equation}
\sigma(dd\rightarrow \mathrm{^4He}\, f_0 \rightarrow 
               \mathrm{^4He}\, \pi^+ \pi^-)= 
               1\div4~ \mathrm{nb}\ .
\label{ddalfaf0}
\end{equation}
  Finally, using Eq.(\ref{eq:dd2hef0}), we get for $|\xi|^2 \simeq 0.05$:
\begin{equation}
  \sigma(dd \rightarrow \mathrm{^4He}\, a_0^0) \simeq 0.05 \div 0.2~
  \mathrm{nb}.
  \label{csest}
\end{equation}

The measurement of reactions (\ref{ddalfaKK}) and (\ref{ddalfaf0})
with cross sections in the sub-nb range are possible at the ANKE
spectrometer. Using a cluster-jet target with Hydrogen as target
material luminosities of $\sim 2.7\cdot10^{31}\ 
\mathrm{cm}^{-2}\mathrm{cm}^{-1}$ have been achieved
\cite{Fedorets_QNP2002}. Assuming that comparable luminosities can be
reached with Deuterium, about 10--40 ($\mathrm{^4He}\, K^+ K^-$)
events can be detected within one week of beam time (based on the
experience of the previous $a_0^+$ beam times). The pions from
reaction (\ref{ddalfaf0}) have broader angular distributions and,
thus, the acceptance of ANKE is about one order of magnitude smaller
which is partially compensated by the larger cross section.

It is planned that within a few years ANKE will be equipped with a
frozen-pellet target \cite{pellet} and a large-acceptance photon
detector \cite{photon}. Since the achievable luminostities then will
be roughly one order of magnitude higher, the isospin-violating
process (\ref{dd}) can be investigated by detecting the decay photons
$\pi^0\rightarrow2\gamma$ and $\eta\rightarrow2\gamma$ in coincidence
with the $\mathrm{^4He}$.  The latter will be again identified and
momentum reconstructed with ANKE.  Based on our cross section estimate
(\ref{csest}) we conclude that a few weeks of beam time will be
sufficient to collect several 100 events.

{\bf Acknowledgment} The authors are grateful for stimulating
discussions with V.~Grishina, C.~Hanhart, J.~Speth and all other
participants of the working group during the workshop.

\end{document}